\def\ion#1#2{#1$\;${\small\rm\@Roman{#2}}\relax}
\documentclass{aa}

\usepackage{graphicx}
\usepackage{txfonts}

\usepackage{natbib}
\begin{document}
   \title{X-ray and UV observations of nova V598 Puppis between 147 and 255 days after outburst}
   \author{K.L. Page\inst{1}, J.P. Osborne\inst{1}, A.M. Read,\inst{1} P.A. Evans\inst{1}, J.-U. Ness\inst{2}, A.P. Beardmore\inst{1}, M. Bode\inst{3}, G.J. Schwarz\inst{4} \& S. Starrfield\inst{5}
          }

   \offprints{K.L. Page}

   \institute{Dept. of Physics and Astronomy, University of Leicester, Leicester, LE1 7RH, UK\\
\email kpa@star.le.ac.uk
\and XMM-Newton Science Operations Centre, ESAC, Apartado 78, 28691 Villanueva de la Ca{\~n}ada, Madrid, Spain
\and Astrophysics Research Institute, Liverpool John Moores University, Birkenhead, CH41 1LD, UK
\and Department of Geology and Astronomy, West Chester University, West Chester, PA 19383, USA
\and School of Earth and Space Exploration, Arizona State University, Tempe,
AZ 85287-1404, USA}

   \date{Received ; accepted }

   \abstract{}{The launch of Swift has allowed many more novae to be observed regularly over the X-ray band. Such X-ray observations of novae can reveal ejecta shocks and the nuclear burning white dwarf, allowing estimates to be made of the ejecta velocity. }{We analyse XMM-Newton and Swift X-ray and UV observations of the nova V598~Pup, which was initially discovered in the XMM-Newton slew survey. These data were obtained between 147 and 255 days after the nova outburst, and are compared with the earlier, brighter slew detection.}{The X-ray spectrum consists of a super-soft source, with the soft emission becoming hotter and much fainter between days $\sim$~147 and $\sim$~172 after the outburst, and a more slowly declining optically thin component, formed by shocks with kT~$\sim$~200--800~eV (corresponding to velocities of 400--800~km~s$^{-1}$). The main super-soft phase had a duration of less than 130~days. The Reflection Grating Spectrometer data show evidence of emission lines consistent with optically thin emission of kT~$\sim$~100~eV and place a limit on the density of the surrounding medium of log(n$_{\rm e}$/cm$^{-3}$)~$<$~10.4 at the 90\% level. The UV emission is variable over short timescales and fades by at least one~magnitude (at $\lambda$~$\sim$~2246--2600~$\AA$) between days 169 and 255.}{}

   \titlerunning{V598 Pup}
   \authorrunning{K.L. Page} 
  \maketitle
%

\section{Introduction}

Novae are cataclysmic explosions which occur in interacting binary systems consisting of a white dwarf (WD) and a lower mass secondary star. Mass is accreted onto the surface of the WD until the pressure and temperature at the base of the envelope reach critical values, at which point a thermonuclear runaway is triggered; see Bode \& Evans (2008) for a review. Any X-ray emission from the nuclear burning source is initially obscured by the ejected material; however, as this envelope expands and becomes optically thin, nuclear burning on the surface of the WD becomes visible. This emission peaks in the soft X-ray band and is known as the Super-Soft Source (SSS) state (Krautter 2008). 

V598~Pup was discovered by Read et al. (2007) in the XMM-Newton slew survey (Saxton et al. 2008) on 9th October 2007 as a transient X-ray source (designated XMMSL1~J070542.7$-$381442) and identified as a nova by Torres et al. (2007).  With a peak magnitude of m$_{\rm v}$~$\la$~4 on 2007 June 5.968 (Pojmanski et al. 2007), V598~Pup is the brightest optical nova seen since V382~Vel and V1494~Aql, both of which were discovered in 1999. V1280~Sco was also found to have a similar peak magnitude (m$_{\rm v}$~$\sim$~3.8) during its outburst in February 2007, just a few months before V598~Pup. Read et al. (2008) estimate the distance of V598~Pup to be $\sim$~3~kpc.

A pointed XMM-Newton observation was performed on 30th October 2007, followed by a series of Swift observations between 21st November 2007 and 15th February 2008; details are given in Table~\ref{obs}. The results of the initial X-ray slew detection were presented by Read et al. (2008), together with optical data obtained by the Magellan Clay telescope and the All Sky Automated Survey (Pojmanski 2002). This paper discusses the later XMM-Newton and Swift X-ray and UV observations.

\begin{table}
\begin{center}
\renewcommand{\arraystretch}{2.0}
\begin{tabular}{cccc} \hline
Satellite & Obs. ID & Day since  & Exp. Time (ks)\\ 
 & & Optical Peak\\
\hline
XMM & 0510010901 & 147 & 5.2 (MOS); 5.1 (pn)\\

Swift & 00031025001 & 169 & 3.8\\
Swift & 00031025003 & 176 & 0.4\\
Swift & 00031025004 & 176 & 7.6 \\
Swift & 00031025005 & 239 & 6.1\\
Swift & 00031025007 & 248 & 4.3\\
Swift & 00031025008 & 255 & 6.4\\

\end{tabular} 
\caption{Log of the observations of V598~Pup presented in this paper, listing the satellite, observation ID number, day of observation and exposure time.}

\label{obs}
\end{center}
\end{table}

\section{Observations and Results}

\subsection{X-ray}

The XMM-Newton and Swift data were processed following the standard methods, using the most up-to-date calibration files available at the time (XMM {\sc sas} 8.0 and Swift 3.2, corresponding to HEASoft release 6.6.3). Spectra and light-curves were extracted using a circular region of radius 20~arcsec centred on the source and a larger background region offset from the nova. Event patterns/grades 0--4 were used for pn, while 0--12 were used for MOS and the Swift observations. Spectra were grouped to have a minimum of one count per bin to allow Cash statistic fitting in {\sc xspec}\footnote{http://heasarc.gsfc.nasa.gov/docs/xanadu/xspec/manual/\\XSappendixCash.html} since this method is found to be less biased than $\chi^2$ (see, e.g., Humphrey et al. 2009).

\begin{figure}
   \centering
   \includegraphics[angle=-90,width=9cm]{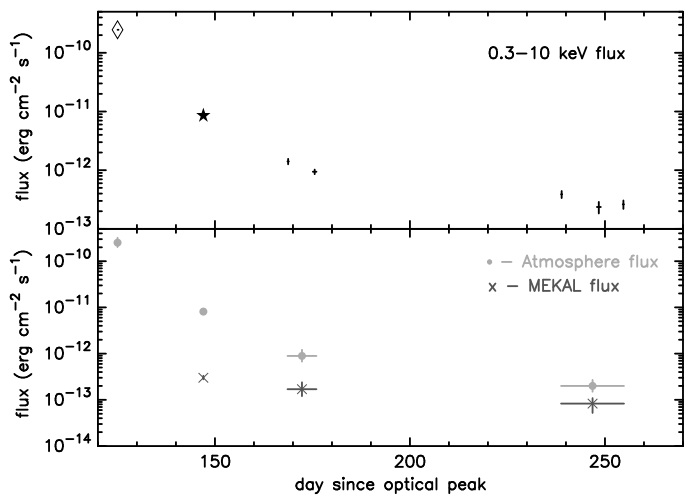}
   \vspace{0.6cm}
   \caption{The top panel shows the 0.3--10~keV unabsorbed flux light-curve of V598~Pup; the diamond and star symbols mark the XMM-Newton slew and pointed data respectively; later observations were by Swift. 
The lower panel plots the flux contributions from the soft (WD atmosphere model) and harder (optically thin) components. The fading of the optically thin component flux can be parameterised by a power-law of slope $\alpha$~$\sim$~2.7,  Before $\sim$~day~161, the soft flux fades rapidly ($\alpha$~$\sim$~21); at later times, the decline is more gradual, with $\alpha$~$\sim$~4. See Sect.~\ref{disc} for details.}         
\label{flux}
   \end{figure}

The top panel of Fig.~\ref{flux} shows the unabsorbed X-ray flux light-curve (over 0.3--10~keV) which follows a fading trend; the source count rates were converted to fluxes using the best fit models at each epoch (see Table~\ref{BBfits} and below). The first measurement is from the XMM-Newton slew data, while the second is from the pointed observation. Here, just the MOS 2 value from the European Photon Imaging Camera (EPIC; Str{\" u}der et al. 2001; Turner et al. 2001) instruments has been plotted for clarity. The following five data points are from the Swift X-ray Telescope (XRT; Burrows et al. 2005).

Spectra were extracted for three time intervals: the XMM-Newton pointed observation taken 147 days after the optical peak, the Swift data taken between 21st and 28th November 2007 (169--176~days after outburst) and the Swift data between 30th January and 15th February 2008 (239--255~days). These spectra were fitted with a combination of either a white dwarf atmosphere model (MacDonald \& Vennes 1991, as used by Balman et al. 1998 for V1974~Cyg) or a blackbody (BB), together with optically thin plasma ({\sc xspec}'s {\sc mekal}; Mewe et al. 1985) components; the {\sc xspec} T{\" u}bingen-Boulder absorption model ({\sc tbabs}) was used with the Wilms abundances (Wilms et al. 2000). See Sect.~\ref{disc} for a comparison of the BB and local thermal equilibrium (LTE) atmosphere models.
It was found that the absorbing column was poorly constrained, but consistent with
 the value  determined by Read et al. (2008) when fitting the slew survey spectrum from a time when the nova was much brighter in X-rays, so their value of N$_{\rm H}$~=~4.8~$\times$~10$^{20}$~cm$^{-2}$ was adopted. 

The results of the atmosphere fits are given in Table~\ref{BBfits}; the corresponding BB temperatures are 32~$\pm$~2, 57$^{+18}_{-30}$ and 82$^{+26}_{-22}$~eV respectively, showing the same trend of increasing temperature between days 147 and 172 after outburst. The fluxes from the atmosphere and optically thin components are plotted in the lower panel of Fig.~\ref{flux}; the BB flux measurements are consistent with these numbers, as expected (see Sect.~\ref{disc}).
The measurement of the soft flux from the slew data has also been plotted. 

\begin{figure}
   \centering
   \includegraphics[angle=-90,width=9cm]{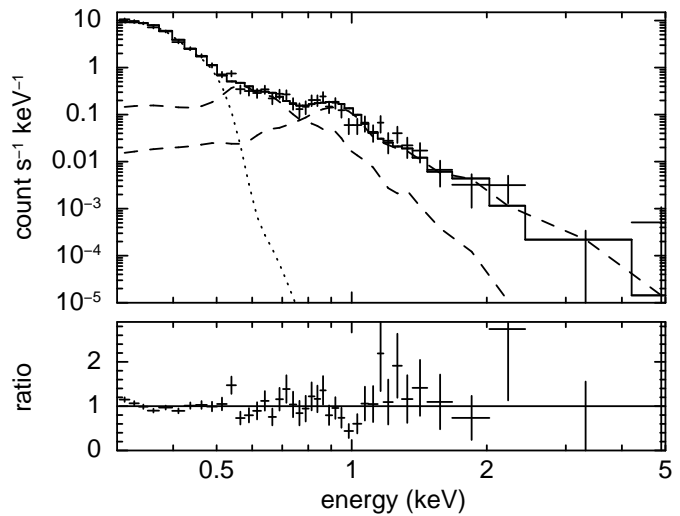}
   \vspace{0.6cm}
   \caption{The pn spectrum of V598~Pup from the pointed XMM-Newton observation. The dotted line indicates the WD atmosphere component (dominating below $\sim$~0.6~keV), while the dashed lines show the optically thin Mekals (at higher energies).}         
\label{spec}
   \end{figure}

Figure~\ref{spec} shows the pointed EPIC-pn spectrum, demonstrating where each of the optically thick (atmosphere) and optically thin components are dominant and showing the clear detection of flux above 0.6~keV. For this observation, two optically thin components are required to achieve a good fit, while the Swift data only required a single one; this difference is probably due to the source being fainter at the time of the Swift observations and does not reflect a real change in spectral components. Indeed, simulating a Swift spectrum (using the {\sc xspec} command fakeit) with two optically thin components of the temperatures required for the EPIC data shows that the result can be well-fitted with just a single temperature component.

\begin{table*}
\begin{center}
\renewcommand{\arraystretch}{2.0}
\begin{tabular}{ccccccccc} \hline
Satellite & Day since & No. of counts & Atmos. kT & Atmos. flux$^{a}$ &  {\sc mekal} kT & {\sc mekal} kT & {\sc mekal} flux$^{a}$ & C-stat/dof\\ 
& Optical peak & in spectrum  & (eV) & (erg cm$^{-2}$ s$^{-1}$)&  (eV)& (eV) & (erg cm$^{-2}$ s$^{-1}$) \\
\hline
XMM & 147 & MOS1: 653 & 35~$\pm$~1$^{b}$ & (8.1~$\pm$~0.3)~$\times$~10$^{-12}$  & 187$^{+14}_{-12}$ & 788$^{+92}_{-86}$ & (3.0~$\pm$~0.4)~$\times$~10$^{-13}$&  383/342 \\
 & & MOS2: 641\\ 
 & & pn: 3458\\
Swift & 172 & 117 & 63$^{+5}_{-4}$ & 8.9$^{+3.2}_{-2.2}$~$\times$~10$^{-13}$&  636$^{+218}_{-335}$ & --& 1.7$^{+0.7}_{-0.5}$~$\times$~10$^{-13}$& 76/56 \\
Swift & 247 & 88 & 68$^{+>18}_{-5}$& 2.0$^{+0.7}_{-0.5}$~$\times$~10$^{-13}$  & 608$^{+153}_{-144}$ & -- & 8.3$^{+2.6}_{-3.1}$~$\times$~10$^{-14}$& 39/54 \\

\end{tabular} 
\caption{WD atmosphere model plus {\sc mekal} fits to V598~Pup spectra at three epochs; the day since peak corresponds to the mid-time of the spectrum. The absorbing column was fixed at 4.8~$\times$~10$^{20}$~cm$^{-2}$ (Read et al. 2008). The XMM-Newton EPIC MOS1, MOS2 and pn spectra were fitted simultaneously, with a free constant of normalisation between them: with the pn constant fixed to unity, MOS 1 and MOS 2 have constants of 0.91~$\pm$~0.08 and 0.96~$\pm$~0.08 respectively. The temperature of the WD atmosphere component for the day 247 observation is constrained by the upper limit allowed by the model.}
$^{a}$ Unabsorbed, over 0.3--10~keV\\
$^{b}$ A systematic error of up to 25\% may be required in addition; see text for details.
\label{BBfits}
\end{center}
\end{table*}

\begin{figure}
   \centering
   \includegraphics[angle=-90,width=9cm]{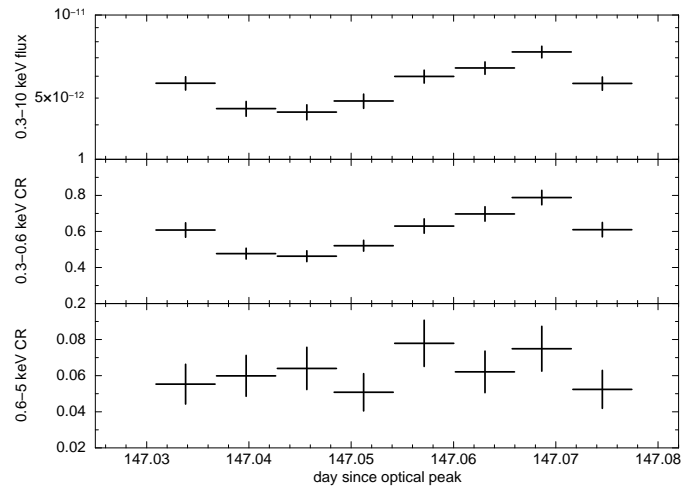}
   \vspace{0.6cm}
   \caption{The top panel panel shows the V598~Pup XMM-pn flux light-curve (0.3--10~keV, unabsorbed in erg~cm$^{-2}$~s$^{-1}$), while the middle and bottom panels show the soft (0.3--0.6~keV) and hard (0.6--5~keV) count-rate light-curves respectively, in units of count~s$^{-1}$. Note the different scales of the lower two panels.}         
\label{xmmlc}
   \end{figure}

Figure~\ref{xmmlc} also focuses on the XMM-Newton data from the pointed observation, showing the EPIC-pn light-curve in 500~s bins; a clear variability of the soft flux is present. The MOS instruments show similar results. The first 1.5~ks of the observation were excluded because of high background seen by all the EPIC instruments. The boundary between the soft and hard bands was chosen to be 0.6~keV since this is the point at which the optically thin emission becomes stronger than the soft component (see Fig.~\ref{spec}). 

The soft-band light-curve is inconsistent with a constant value, with a reduced $\chi^2$ of $\sim$~10, whereas the hard band has an acceptable $\chi^2_\nu$~=~0.7. Fitting the soft data with a sine wave simply to characterise the variation (no periodic modulation is inferred) gives $\chi^2_\nu$~=~0.95. The amplitude of the modulation is (15~$\pm$~3)\% over the 0.3--0.6~keV band; applying this same model to the hard band (keeping the phase fixed) provides a 90\% confidence upper limit on the modulation of 2\% over 0.6--5~keV. These results demonstrate that the variability seen is almost entirely caused by the soft spectral component.  Such variability is a common feature of novae supersoft emission (e.g., Orio et al. 2002; Osborne et al. 2006a; Drake et al. 2008; Page et al. 2009).

\begin{figure}
   \centering
   \includegraphics[angle=-90,width=9cm]{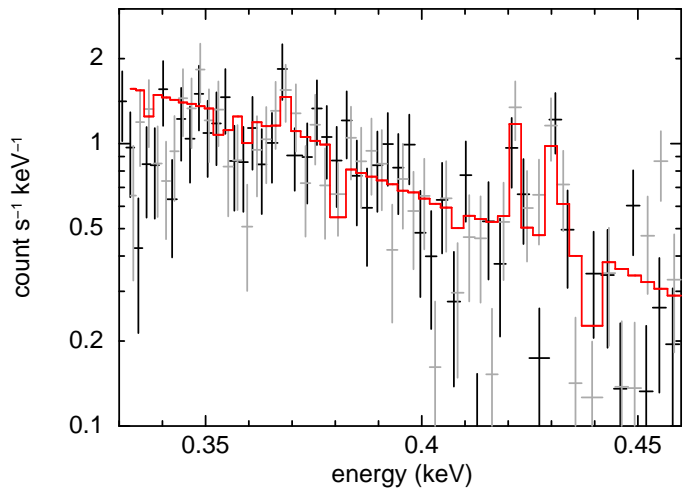}
   \vspace{0.6cm}
   \caption{The RGS spectra of V598~Pup (RGS1 is shown in black, RGS2 in grey), fitted with a atmosphere model and individual emission features from C VI and N VI. The apparent absorption features in the model are due to the detector.}         
\label{rgs}
   \end{figure}

Figure~\ref{rgs} plots the Reflection Grating Spectrometer (RGS; den Herder et al. 2001) data obtained on day 147. The spectra clearly show narrow emission features, which can be modelled by an optically thin plasma of kT~=~120$^{+22}_{-20}$~eV together with a soft, optically thick component of 32~$\pm$~1~eV to fit the smooth continuum. The optically thin temperature inferred from the RGS data is cooler than that determined from the EPIC data which suggests that the emission is not isothermal. The RGS spectra are fitted over a much narrower and lower-energy bandpass than the MOS and pn, so will be less sensitive to the higher temperatures seen by EPIC.

Instead of using a generic optically-thin model component, individual features can be identified in the RGS spectrum and
corresponding emission lines superimposed on the optically thick continuum. The strongest lines (all $>$3$\sigma$ detections) are at 368.5$^{+0.4}_{-0.5}$~eV (equivalent width, EW, of 1.7~$\pm$~1.2~eV, line normalisation of 1.7$^{+1.2}_{-0.7}$~$\times$~10$^{-4}$ ph~cm$^{-2}$~s$^{-1}$; likely due to C VI emission), 421.4$^{+0.5}_{-0.4}$~eV (EW~=~5.1~$\pm$~2.5~eV, line normalisation of 6.9$^{+6.2}_{-4.3}$~$\times$~10$^{-5}$ ph~cm$^{-2}$~s$^{-1}$; N VI forbidden line) and 430.5$^{+0.4}_{-0.5}$~eV (EW~=~4.4~$\pm$~2.2~eV, line normalisation of 1.1$^{+0.6}_{-0.5}$~$\times$~10$^{-4}$ ph~cm$^{-2}$~s$^{-1}$; N VI resonance line). The N VI intercombination line would be expected at 426.4~eV, but no significant emission is found at this energy; the 90\% upper limit on the EW is 4~eV ($<$7.7~$\times$~10$^{-5}$  ph~cm$^{-2}$~s$^{-1}$ for the line normalisation). The ratio of the strengths of the forbidden and intercombination lines can be used to place a limit on the density of the medium of log(n$_{\rm e}$/cm$^{-3}$)~$<$~10.4 at the 90\% level (Gabriel \& Jordan 1969, Porquet \& Dubau 2000; Ness et al. 2004). A strong resonance line is an indication of a plasma where collisional processes are significant, confirming the assumptions made in the earlier spectral fits.

\subsection{Ultraviolet}

Swift also observed V598~Pup with the three UV/Optical Telescope (UVOT; Roming et al. 2005) UV filters (uvw1: central wavelength 2600~$\AA$; uvm2: 2246~$\AA$; uvw2: 1928~$\AA$), except on day 176. The source magnitudes were calculated using {\sc uvotmaghist}, with a standard 5~arcsec radius source extraction region (Poole et al. 2008) for the source and a 25~arcsec radius circle for the background.

\begin{figure}
   \centering
   \includegraphics[angle=-90,width=9cm]{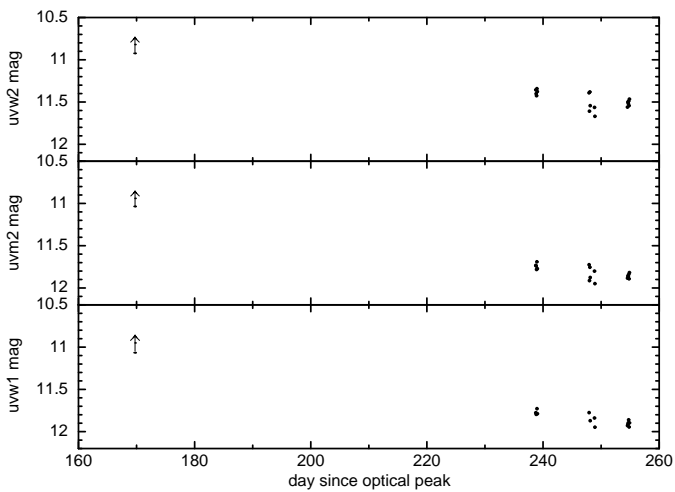}
   \vspace{0.6cm}
   \caption{Swift-UVOT light-curves of V598~Pup in the UV filters. The left-hand panel shows the early, saturated data, where only a lower limit can be placed on the magnitude. The right-hand panel shows the later observations.}         
\label{uvot}
   \end{figure}

During the observation on day 169 the source was bright enough to saturate the detector, so the measurements obtained are not reliable; thus only lower limits can be set on the magnitude, with the conclusion that the UV emission on $\sim$~day~169 was at a magnitude of $\la$~11. Hence, over $\sim$~86 days, the UV emission faded by a magnitude or more in the uvm2 and uvw1 filters, while the emission in the shortest wavelength filter (uvw2) only decreased by around half this value (Fig.~\ref{uvot}). Short-term variability can also be seen on timescales of less than a day. 

\section{Discussion}
\label{disc}

The launch of Swift has led to many novae being observed repeatedly in
the X-ray band (see, e.g., Bode et al. 2006; Osborne et al. 2006b;
Ness et al. 2007, 2008, 2009; Lynch et al. 2008; Bode et al. 2009;
Page et al. 2009), allowing us to follow the changes which occur
during the outbursts. Some novae (e.g. V1494~Aql -- Drake et al. 2003;
V4743~Sgr -- Ness et al. 2003) were found to show relatively rapid
X-ray variability even in the pre-Swift era, but the Swift
observations have enlarged this sample dramatically.  The observations
of V598~Pup were unusual, however, in that the nova was first
discovered in X-rays rather than optically; observations began more
than 100~days after the optical outburst, when the supersoft X-ray
phase was already fading.  Here, we have presented the first detection
of an optically thin component in this nova; the discovery spectrum
presented by Read et al. (2008) could be well fitted by optically
thick emission alone.

Both BB and LTE WD atmosphere
models have been used in the past to model the SSS spectra of novae,
although there are problems with each method. BB models are extremely
simplified, essentially ignoring all absorption processes. They are,
thus, not physical representations of atmospheres. Bolometric
luminosities derived from such fits are overestimated by up to two
orders of magnitude compared to those derived from LTE model
atmospheres, while the temperatures are 20-30\% lower than those determined from model atmosphere fits (e.g., Heise et al. 1994; Krautter et al. 1996). It
has been argued that stellar atmosphere models are more physical
because they account for radiation transport in a self-consistent way
(e.g., Balman et al. 1998) but these models make simplifying
assumptions as well. The extended atmosphere in the late stage of a nova likely requires full non-LTE treatment for a complete parameterisation (Hauschildt 2008) and we are aware of no currently available atmosphere model
which is able to account for the expansion of nova ejecta. Because of the shortcomings of the different models, it seemed instructive to present results from both methods in this paper. We also
note that the band-limited, X-ray flux measured is model-independent
within the uncertainties, providing the model is an acceptable fit to
the data; thus, the flux estimates are approximately the same whether
BB or WD atmosphere models are used.

The underlying (harder) flux from the optically thin plasma fades only slightly over time, whereas the soft flux drops by almost two orders of magnitude between days 147 and 247 (more than three orders of magnitude from day 125 to 247), leading to most of the flux decay seen in the top panel of Fig.~\ref{flux}. Parameterising the decay with a power-law, the higher energy, optically thin flux follows a slope of $\alpha$~=~2.7$^{+1.8}_{-1.3}$ [flux~$\propto$~(t-t$_0$)$^{\alpha}$, where t$_0$ is taken to be the time of the optical peak and t is in days]. The softer model atmosphere flux, however, does not follow a simple power-law decline, with the decay being more rapid before the Swift observations begin.  Such a change in the rate of X-ray decline has been previously seen in other novae, e.g. RS~Oph (Osborne et al. 2006b, 2009), V2491~Cyg (Page et al. 2009) and V382 Vel (Burwitz et al. 2002).  With only four flux points and four free parameters in the model, the decay slopes for the soft flux can only be approximated, but the decay indices are $\alpha_{\rm 1}$~$\sim$~21$^{+2}_{-4}$ between the XMM and Swift observations  (until $\sim$~day~161) and $\alpha_{\rm 2}$~$\sim$~4~$\pm$~2 at later times; this is consistent with the late-time decay of the soft flux in V2491~Cyg, for which $\alpha_{\rm soft}$~=~3.08~$\pm$~0.04 is measured (Page et al. 2009).

There is an apparent increase in temperature of the atmosphere component between the XMM and Swift observations;  such an increase is expected as the optical depth of the atmosphere decreases, revealing the hotter regions at smaller radii.
 The temperature on day 147 is consistent with that measured for the slew spectrum by Read et al. (2008).
The atmosphere model we use has an upper boundary of 86~eV for the temperature, which does not allow the calculation of the upper uncertainty for the Swift data obtained around day 247. This is a problem which has also been found previously (e.g., for V2491~Cyg; Page et al. 2009).

We note that both XMM-Newton and Swift have been cross-calibrated on the soft neutron star source RXJ~1856.4$-$3754 (which is fitted with a BB of 63~eV, following Beuermann et al. 2006; see Godet et al. 2009), so the observed temperature increase between days 147 and 172 is not likely to be an artifact of the inter-instrument calibration. However, the uncertainty on the XMM-Newton data given in Table~\ref{BBfits} may be underestimated for such a soft source (S. Sembay, private communication); a systematic error of up to 25\% on the temperature may be applicable at such low values. The temperatures of the soft emission detected in the Swift datasets (Table~\ref{BBfits}) are comparable to those predicted for the decline phase of a nova (Prialnik 1986) and the flux of the soft component is decreasing rapidly, as described above. We can, therefore, place a firm upper limit of $\sim$~130~days on the duration of the constant X-ray luminosity phase, with the end probably being no later than the time of the XMM slew observation. With the available data, it is not possible to determine whether the XMM-Newton slew data were obtained during the maximum X-ray luminosity phase, or if the source was already declining. There are few novae with well-measured durations for the constant X-ray luminosity phase -- other examples include V1974~Cyg ($\sim$~256~days -- Krautter et al. 1996; Balman et al. 1998), RS~Oph ($\sim$~12 days -- Osborne et al. 2009) and V2491~Cyg ($\sim$~2 days -- Page et al. 2009), with V723~Cas (Ness et al. 2008) remaining an SSS more than 12~years after its nova outburst; the observations of V598~Pup are a useful addition to the sample.


Although we simultaneously fit optically thick and thin components, it could be that the soft emission which we have parameterised by an optically thick WD atmosphere model is contaminated in the Swift-XRT spectrum with a blend of emission lines from the optically thin component(s). However, we note that the EPIC CCD spectrum is best-fitted with an optically thick component of a temperature consistent with that derived from the simultaneous RGS data. With this in mind, we see no strong reason to suspect that the temperature derived for the optically thick emission from the Swift-XRT is significantly affected by emission lines from optically thin gas, although the flux of the source is lower at this time.

Both the atmosphere and BB models allow the estimation of the size of the emitting region. Using the stellar atmosphere model, this radius is determined to be $\sim$~1~$\times$~10$^{8}$, 6~$\times$~10$^{6}$ and 3~$\times$~10$^{6}$~cm during the XMM-Newton and Swift observations respectively (2~$\times$~10$^{8}$, 6~$\times$~10$^{6}$ and 1~$\times$~10$^{6}$~cm for the BB model), taking the distance of V598~Pup to be 3~kpc. These numbers are smaller than the typical radius of a WD ($\sim$~10$^9$~cm), possibly exposing the limitations of the models used; such a result was also found for V2491~Cyg (Page et al. 2009).

The shortest wavelength (uvw2) UV emission is noticeably brighter than that seen in the other UV filters and fades by a smaller amount ($\ga$~0.5~mag, compared to $\ga$~1~mag in the uvm2 and uvw1 filters; these are lower limits because of the saturation of the early time data). The difference between uvw2 and the other UV filters could be explained by dominant line emission -- as the ejecta cool, low ionisation lines start to reappear.  For example, the C III] line at 1909~$\AA$ can be very strong (Schwarz 2002; Cassatella et al. 2005; Schwarz et al. 2007) and would appear predominantly in the uvw2 filter.
It is not possible to separate the continuum and line emission without spectroscopic observations.

There is no strong correlation between the variability seen in the X-ray and UV over short periods of time (i.e., within a single snapshot). Considering the final observation ($\sim$~day 254.5--254.9; this was the longest of the UVOT observations), the Spearman Rank test gives a $\la$~95\% probability of the X-ray and UV bands being anti-correlated. Novae observed by Swift have shown a variety of behaviours when comparing the X-ray and UV emission: some, such as V598~Pup and V2491~Cyg (Page et al. 2009) show little discernible correlation; V458~Vul shows anti-correlative behaviour (Drake et al. 2008; Ness et al. 2009), while the X-ray and UV data obtained for CSS~081007:030559+054715 vary approximately in phase (Beardmore et al. 2008; paper in prep). 

However, the variability seen in these other novae has been mainly when there was still significant SSS emission. In the case of V598~Pup, the useful UV data were collected when the main supersoft phase had ended. The (optically thin) X-ray emission at this time is likely to be due to shocks caused by components within the ejecta moving at different velocities. The temperatures found for these shocks in V598~Pup ($\sim$~200--800~eV) are somewhat lower than those found for V2491~Cyg (Page et al. 2009) or RS~Oph (Bode et al. 2006), though more similar to V1974~Cyg (Balman et al. 1998), for example. The corresponding shock velocities (due to differential motion within the ejecta) are between $\sim$~400 and 800~km~s$^{-1}$.
Using the distance estimate of 3~kpc from Read et al. (2008), the bolometric luminosity of this hard emission is $\sim$~(0.5--1)~$\times$~10$^{33}$~erg~s$^{-1}$ over the time of the observations.

Swift observations allow nova outbursts to be followed in detail and are 
continuing to shed light on their X-ray behaviour -- both the shocks seen in the ejecta (e.g., Bode et al. 2006) and the (sometimes highly variable) supersoft emission (e.g., Osborne et al. 2009; Page et al. 2009; Ness et al. 2008). Such analysis is leading to a more detailed understanding of the processes which occur during the evolution of novae.

\begin{acknowledgements}
We thank the Swift PI, Neil Gehrels, and the science and mission operations teams for their support of these observations. 
XMM-Newton is an ESA science mission with instruments and contributions directly funded by ESA member states and NASA in the USA. 
KLP, JPO, AMR, PAE and APB acknowledge support from STFC. SS acknowledges support from Chandra and NSF. KLP thanks DPF for a consultation on Fig.~\ref{uvot}.
\end{acknowledgements}

\end{document}